\documentclass[aps,prl,twocolumn,showpacs,superscriptaddress,groupedaddress]{revtex4-1}
\usepackage{amsmath, amssymb}
\usepackage{graphicx}
\usepackage{mathptmx}

\begin{document}

\title{Schr\"odinger's Cat in an Optical Sideband}
\author{Takahiro Serikawa,$^{1}$ Jun-ichi Yoshikawa,$^{1}$ Shuntaro Takeda,$^{1}$ Hidehiro Yonezawa,$^{2,3}$\\ Timothy C. Ralph,$^{3,4}$ Elanor H. Huntington,$^{3,5}$ and Akira Furusawa$^{1}$}
\email[]{akiraf@ap.t.u-tokyo.ac.jp}
\affiliation{$^{1}$Department of Applied Physics, School of Engineering, The University of Tokyo, 7-3-1 Hongo, Bunkyo-ku, Tokyo, Japan.}
\affiliation{$^{2}$School of Engineering and Information Technology, The University of New South Wales, Canberra, ACT 2600, Australia}
\affiliation{$^{3}$Center for Quantum Computation and Communication Technology, Australian Research Council, Australia.}
\affiliation{$^{4}$School of Mathematics and Physics, University of Queensland, Brisbane, QLD 4072, Australia.}
\affiliation{$^{5}$Research School of Engineering, College of Engineering and Computer Science, Australian National University, Canberra, ACT 2600, Australia.}

\date{\today}

\begin{abstract}
We propose a method to subtract a photon from a double sideband mode of continuous-wave light. The central idea is to use phase modulation as a frequency sideband beamsplitter in the heralding photon subtraction scheme, where a small portion of the sideband mode is downconverted to the carrier frequency to provide a trigger photon. An optical Schr\"odinger's cat state is created by applying the propesed method to a squeezed state at 500MHz sideband, which is generated by an optical parametric oscillator. The Wigner function of the cat state reconstructed from a direct homodyne measurement of the 500\,MHz sideband modes shows the negativity of $W(0,0) = -0.088\pm0.001$ without any loss corrections.
\end{abstract}

\maketitle

Implementation of quantum operations or creation of quantum states on multiplexed photonic modes is a key for universal and scalable photonic quantum information processing (QIP).
Time-division or frequency-division multiplexing provides the means of compact generation and manipulation of numerous quantum states.
Recent demonstrations of large-scale continuous-variable (CV) cluster states \cite{PhysRevA.73.032318} in time \cite{yokoyama2013ultra} and frequency \cite{PhysRevLett.107.030505,PhysRevLett.112.120505} domains are excellent examples of multiplexed quantum optics, though they belong to Gaussian states and transformations.
Employing the cluster states, CV one-way quantum computing model \cite{raussendorf2001one,PhysRevLett.97.110501} offers a framework of QIP, where ancillary non-Gaussian states or measurements are required for its universality \cite{PhysRevLett.82.1784,PhysRevLett.97.110501,PhysRevA.71.042308}.

Photon subtraction \cite{PhysRevA.55.3184,0953-4075-41-13-133001} is a common method to create non-Gaussian states, and has been established on baseband photonic modes. It is a versatile technique and has wide applications, such as quantum noiseless amplification \cite{zavatta2011high}, entanglement enhancement \cite{PhysRevA.86.012328,PhysRevA.87.022313}, or a creation of particle-wave hybrid entanglement \cite{morin2014remote}.
An optical Schr\"odinger's cat (SC) state is a famous example of non-Gaussian states created by means of subtracting a photon from a squeezed vacuum state \cite{ourjoumtsev2006generating,PhysRevLett.97.083604,ourjoumtsev2007generation}. SC states are powerful resources to implement several applications of QIP such as quantum error correction \cite{PhysRevLett.111.120501,ofek2016extending} or quantum computing based on coherent states \cite{PhysRevA.68.042319}.
Incorporating frequency-domain techniques in the photon subtraction scheme will lead to universal and practical quantum operations over multiplexed photonic modes.

High-frequency sideband modes are desirable target for the frequency-division multiplexing, since such modes can be broadband. The bandwidth is practically important, especially when they are combined with the time-domain techniques such as time-bin encoding \cite{PhysRevLett.82.2594,PhysRevA.87.043803} or time-domain cluster state computation \cite{PhysRevLett.97.110501,yokoyama2013ultra}. Here, to access a certain optical mode at high-frequency sideband for photon subtraction, we need to selectively tap off and detect a photon in the target mode. This is a challenging task because sideband modes are sinusoidal wave on an optical beam and higher frequency modes requires higher timing resolution to be addressed.

In this Letter, we propose a method to do photon subtraction in a manner that can be easily extended to creation of multiple non-Gaussian states on high-frequency modes of a single laser beam.
For the basis of the subtraction process, an optical double sideband (DSB) mode, i.e. a balanced superposition of upper and lower sideband modes around a carrier frequency, is employed.
The proposed method is experimentally applied to a squeezed state generated by an optical parametric oscillator (OPO). An SC state is heralded on the 500.6\,MHz DSB mode. 
The bandwidth of the created cat state is about 5\,MHz which is comparable to that of the conventionally demonstrated optical non-Gaussian state generation.
State verification is done by homodyne tomography and the cat state has excellent negativity in the Wigner function. The negativity is directly measured on the high-frequency sideband without loss correction, showing that the quantum non-Gaussianity can be actually used for applications that include measurement and feedforward, such as one-way quantum computing.

A DSB mode is described as $\bigl(e^{i\theta}\hat{a}_\Omega + e^{-i\theta}\hat{a}_{-\Omega}\bigr)/\sqrt{2}$, where $\hat{a}_\Omega$ is an annihilation operator at frequency $\Omega$ around the carrier, and $\theta$ is an arbitrary phase. In time-domain, it has a real, sinusoidal envelope $\cos(\Omega t + \theta)$.
To access DSB modes, phase or amplitude modulation can be used; for example displacement operations has been implemented on DSB modes by a modulator and a beamsplitter. Since DSB modes are apart of the carrier frequency, they are free from the technical noise around the carrier, which enables shot-noise-limited measurement of the field amplitude, leading to, for example, an atomic quantum memory of a DSB light realized by measurement and feedback \cite{julsgaard2004experimental}.
Here, corresponding to two degrees of freedom of $\hat{a}_\Omega$ and $\hat{a}_{-\Omega}$, DSB modes at frequency $\Omega$ are decomposed into two quadrature phase components, namely {\it cos-sideband} $\hat{a}_\Omega^{\cos} = \bigl(\hat{a}_\Omega + \hat{a}_{-\Omega}\bigr)/\sqrt{2}$ and {\it sin-sideband} $\hat{a}_\Omega^{\sin} = \bigl(\hat{a}_\Omega - \hat{a}_{-\Omega}\bigr)/\sqrt{2}i$.
Thus dealing with DSB modes is always a multi-mode problem; photon subtraction should selectively access one of them.

\begin{figure}[t]
 \centering
 \includegraphics{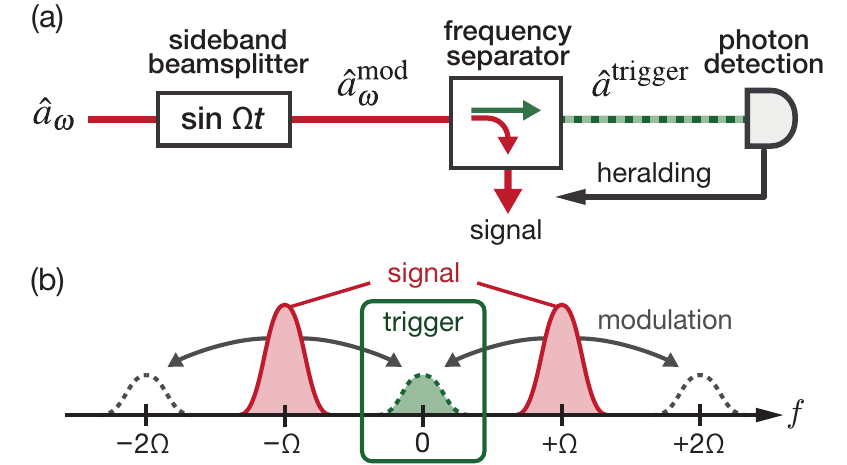}
 \caption{\label{fig:subt} (a) Schematics of photon subtraction from a double sideband. (b) Frequency diagram. A phase modulation with the signal $\sin \Omega t$ is applied to the input light. The cos-sideband component is coupled to the carrier frequency mode, which is initially prepared as a vacuum state. The carrier frequency photon is spatially separated from the sideband signal and guided to the photon detector. The arrival of trigger photon heralds photon subtraction from the sideband.}
\end{figure}

The concept of our method is depicted in Fig. \ref{fig:subt}. A small portion of the signal light at frequency $\Omega$ is downconverted to the carrier frequency by a {\it sideband beamsplitter}. This is realized by a small phase modulation, which transfers an optical component at a given frequency to both upper and lower sidebands \cite{Capmany:10}. 
In the Heisenberg picture, weak frequency-$\Omega$ modulation transforms $\hat{a}_\omega$ as
\begin{align}
\hat{a}^\mathrm{mod}_\omega \sim \sqrt{1-\frac{\beta^2}{2}}\hat{a}_\omega + \frac{\beta}{2} \bigl(e^{i\theta}\hat{a}_{\omega+\Omega} + e^{-i\theta} \hat{a}_{\omega-\Omega}\bigr),
\label{eq:mod}
\end{align}
where $\beta \ll 1$ expresses the modulation depth and $\theta$ is determined by the modulation phase.
This creates a superposition of upper and lower sidebands at 0\,Hz with the simple setup, which is challenging if we use a straightforward implementation of frequency-domain interaction, i.e frequency separation, shift and mixing.
The frequency separator passes the carrier frequency component on to the trigger mode. Subsequent photon detection at the carrier frequency herald photon subtraction events, which can be expressed as conditioning by a single-photon state of the trigger mode as
\begin{align}
 {}_\mathrm{trigger}\langle1| \sim {}_\mathrm{sig}\langle0|\left[\hat{a}_0 + \frac{\beta}{\sqrt{2}} \frac{e^{i\theta} \hat{a}_{\Omega} + e^{-i\theta}\hat{a}_{-\Omega}}{\sqrt{2}}\right],
 \label{eq:subtraction}
\end{align}
where the creation operator of the trigger mode is reduced to the signal modes by Eq. (\ref{eq:mod}). Since the initial state of carrier frequency mode $\hat{a}_0$ is assumed to be vacuum, the conditioning with Eq. (\ref{eq:subtraction}) results in photon subtraction on the DSB mode with the phase $\theta$. A strong advantage of our method is that we can select the cos-sideband, or a DSB mode with any phase, since $\theta$ is controllable by tuning the modulation phase. Note that the effect of the finite linewidth of the separator is ignored here. Actually, a photon is subtracted from a wavepacket as conventional baseband subtraction methods; see Supplement Material for a further formulation.

A significant advantage of the DSB basis is that highly-multiplexed, potentially over thousands of, squeezed vacuum states in DSB modes are  available by a continuously-pumped optical parametric oscillator (OPO) \cite{PhysRevLett.112.120505,PhysRevA.73.013817}.
The photon-pair generation process of a degenerate OPO is described by $\exp\bigl(\int_0^\infty d\omega\, r(\omega)\, \hat{a}^\dagger_\omega\hat{a}^\dagger_{-\omega} - \mathrm{h.c.}\bigr)$ where $r(\omega)$ denotes the squeezing spectrum, which has comb-like shape corresponding to the resonances of the OPO.
With the DSB basis, this is reinterpreted as two photon creation / annihilation process of each DSB mode since $\hat{a}^\dagger_\omega\hat{a}^\dagger_{-\omega} =  \bigl[(\hat{a}^{{\cos}\dagger}_\omega)^2 + (\hat{a}^{{\sin}\dagger}_\omega)^2\bigr] /2$. Thus we have independent squeezed states on both sin- and cos-sideband modes, which include even thousands of frequency combs \cite{wang2014engineering} and can be used for the resource of non-Gaussian state generation.

\begin{figure*}[t]
 \centering
 \includegraphics{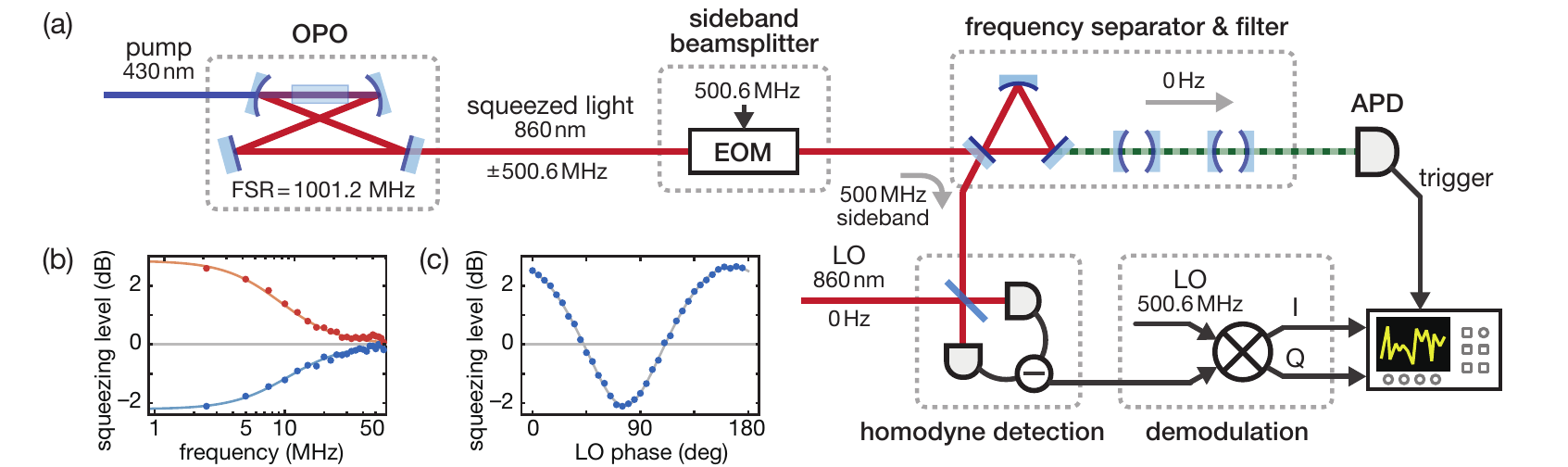}
 \caption{\label{fig:setup} (a) Schematic of the experiment. I and Q denote in-phase and quadrature component of the homodyne signal. (b) Squeezing / anti-squeezing spectrum around the 500.6\,MHz sideband. The power spectrum is calculated by fast Fourier transformation of the homodyne detection. This is an average of 8000 traces of 400\,ns period, and normalized by the shotnoise power. (c) Phase scan plot of the low-frequency squeezing level. Theoretical curves are also shown.}
\end{figure*}

Figure \ref{fig:setup}(a) shows the experimental setup. An SC state is created by subtracting a photon from a squeezed vacuum state at a DSB mode, which is prepared by an OPO.
We carefully identify the free spectral range (FSR) of the OPO at $2\Omega = 1001.2\,\mathrm{MHz}$ to determine the sideband frequency  $\Omega = 500.6\,\mathrm{MHz}$.
Our OPO is resonant at $(2n+1)\Omega,\ n \in \mathbb{Z}$ with the linewidth of 10\,MHz and the carrier frequency mode is kept vacuum. 
Since the squeezing operation of the OPO can be factorized in the sin- and cos-sideband modes, the squeezed state is separable in the DSB basis. When we only look at carrier frequency and the first resonance at $\Omega$, the output of the OPO is expressed as
\begin{align}
 |\Psi_0\rangle = |0\rangle_\mathrm{0} \otimes \hat{S}_r|0\rangle_{\cos} \otimes\hat{S}_r|0\rangle_{\sin},
 \label{eq:initial}
\end{align}
where $|0\rangle_\mathrm{0}$ is a vacuum state of the carrier mode $\hat{a}_0$ and $\hat{S}_r|0\rangle_{\cos, \sin}$ are squeezed states of $\hat{a}_\Omega^{\cos}$ and $\hat{a}_\Omega^{\sin}$, respectively. For the simplicity, we omit the multi-mode description of the continuous-wave squeezed light here; again, see Supplemental Material.

In order to apply phase modulation at 500.6\,MHz without inducing decoherence, we use a bulk electro-optic modulator (EOM) that has low-optical loss below 0.5\%. The transfer efficiency $\beta^2$ is set at 0.040. By adjusting the phase of the driving signal of the EOM, cos-sideband mode is selectively downconverted to the carrier, i.e. $\theta$ in Eq. (\ref{eq:mod}) is set at zero. The frequency separator consists of three optical cavities, and extracts the trigger photon component at the carrier frequency with about 5\,MHz of bandwidth, while rejecting all the higher frequency resonances of the OPO over several hundred GHz. The clicks of the avalanche photodiode (APD) provide the trigger signal for photon subtraction. Applying Eq. (\ref{eq:subtraction}) on Eq. (\ref{eq:initial}) yields an SC state in the cos-sideband mode, while the sin-sideband mode remains as a squeezed vacuum state:
\begin{align}
 |\Psi_\mathrm{cat}\rangle \propto {}_\mathrm{trigger}\langle1|\Psi_0\rangle \propto \hat{a}^\mathrm{cos}_\Omega\hat{S}_r|0\rangle_{\cos} \otimes\hat{S}_r|0\rangle_{\sin}.
\end{align}
The SC state actually has a wavepacket-like envelope $\xi(t)$ and is generated in a sideband wavepacket $\cos\Omega t\, \xi(t-\tau)$ around the trigger time $\tau$. The shape of the envelope is determined by the frequency characteristics of the squeezed state and the transmission spectrum of the frequency separator, which are tunable parameters and in principle can be matched to external devices such as optical memories.

The quadrature distributions of the sin- and cos-sideband modes are measured by homodyne detection with a continuous-wave optical local oscillator (LO) at the carrier frequency.
83\% of effective detection efficiency is realized at 500\,MHz by a low-loss, low-noise resonant homodyne detector \cite{doi:10.1063/1.5029859}.
The two DSBs are electrically resolved by a demodulator with a pre-defined electrical LO at frequency $\Omega$, giving cos- and sin-sideband quadrature as in-phase and quadrature-phase output.

Figure \ref{fig:setup}(b) shows the squeezing spectrum at the 500.6\,MHz sideband calculated from the quadrature-phase component (sin-sideband mode) of the homodyne detection. We obtain 2.2\,dB of squeezing and the total efficiency of sin-sideband is estimated at $\eta^{\sin}=0.70$. Figure \ref{fig:setup}(c) is the phase scan plot of the squeezing level averaged within DC-5\,MHz. The squeezing phase is estimated at 66 degrees. The phase of the squeezed state can be easily changed by adjusting the pump phase locking.

For the tomography of the cat state, the in-phase (cos-sideband) and quadrature (sin-sideband) signals are simultaneously digitized with the trigger signals. 8,000 samples of quadrature signals for each 36 equally partitioned optical phases are collected.

\begin{figure}[b]
 \centering
 \includegraphics{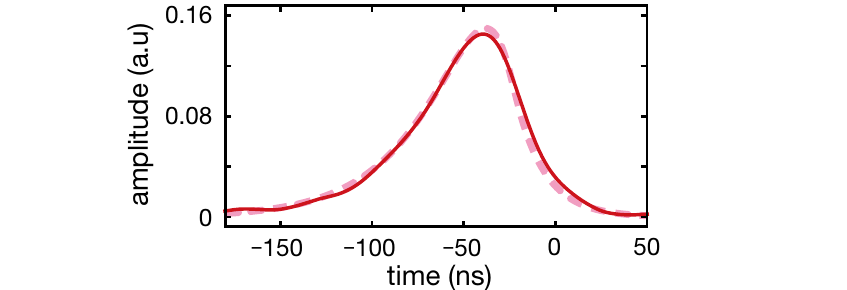}
 \caption{\label{fig:wavepacket} Estimated envelope function $\xi(t)$ of the sideband wavepacket of the subtracted mode. The time origin is placed at the trigger time. The dashed line shows the theoretical curve.}
\end{figure}

The envelope function $\xi(t)$ of the SC states are identified by independent component analysis \cite{comon1994independent} of the demodulated cos-sideband waveforms and shown in Fig. \ref{fig:wavepacket}. The estimated $\xi(t)$ has about 5\,MHz of bandwidth, and well matches the theoretical curve, which is obtained as a convolution of the correlation function of the OPO and the impulse response of the trigger line filters. Since the bandwidth of the trigger line filters is narrower than that of the OPO, the envelope function resembles the single-sided decay function of the filter's response. The quadrature of the wavepacket of an SC state is given by a weighted integration of in-phase signal with $\xi(t)$, which is realized by a digital filter and the impulse response of the homodyne detector (see Supplemental Material). In order to discuss the sideband-selectivity of our method, we also extract the quadrature of the sin-sideband wavepacket that has the same envelope as the photon-subtracted state.

\begin{figure}[t]
 \centering
 \includegraphics{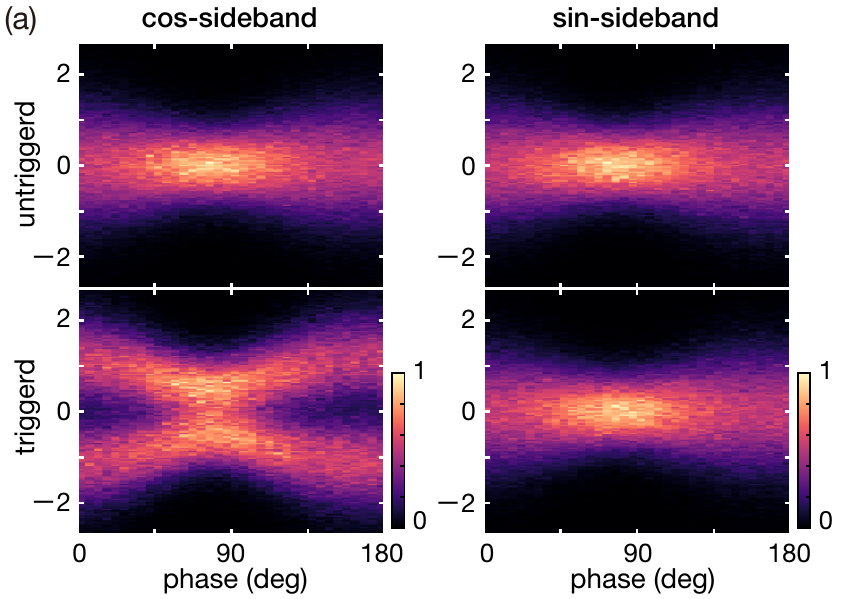}
 \includegraphics{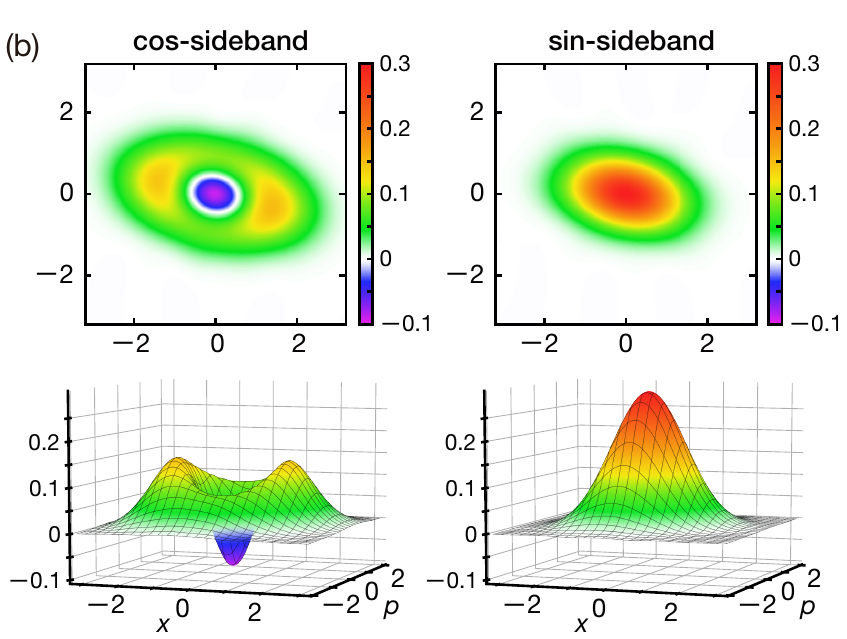}
 \caption{\label{fig:result} (a) Quadrature distributions of 36 phase slices ($\hbar = 1$). Upper row: recorded regardless of trigger. Lower row: triggered. (b) Reconstructed Wigner functions. This is directly observed data and no analytical corrections for experimental imperfections are applied.}
\end{figure}

The quadrature distributions of cos- and sin-sideband wavepacket modes show the effect of subtraction (Fig. \ref{fig:result}(a)), where only the cos-sideband state is reshaped by the conditions of the triggers. The non-classical nature of the generated state is confirmed by the negativity of the Wigner function obtained by maximum-likelihood estimation \cite{lvovsky2004iterative} (Fig. \ref{fig:result}(b)). The cos-sideband state shows $W_{\cos}(0,0) = -0.088 \pm 0.001$ ($\hbar = 1$) without loss correction, which is to be compared with the negative peak of the pure SC states $W_\mathrm{cat}(0,0) = -1/\pi$. The fidelity of the cos-sideband state to the best-fit minus cat state $|\Psi\rangle = \mathcal{N}\bigl[|\alpha\rangle - |-\alpha\rangle\bigr]$, with the coherent state amplitude $\alpha = 0.88 -0.19i$, is 64\%. Both optical losses and contamination from the sin-sideband contribute to $W(0, 0)$ as a mixture of plus value $1/\pi$. In this sense, when the estimated total efficiency $\eta^{\cos} = 0.68$ (see Supplemental Material) is considered, we expect $W_{\cos}(0,0) = -0.114$. To fit the actual value of $W_\mathrm{cat}(0,0)$, 4\% of mixture of background squeezed state is presumed where the fake clicks of the APD and the impurity from the inherent mode-mismatch of photon subtraction \cite{PhysRevA.96.052304} contributes 0.8\% and 3.0\% to it respectively. Thus the upper bound of the mixture of the sin-sideband component is estimated below 1\%. The sin-sideband mode has 99.9\% fidelity to the lossy squeezed state since it is untouched by the subtraction. There is no extra factors that limit the purity in our method than the conventional photon subtraction; the major imperfection is the detection efficiency which is relatively low compared to the baseband experiments \cite{Asavanant:17}.
Our work can be compared with the recent works by Averchenko {\it et al.} \cite{PhysRevA.89.063808} and Ra {\it et al.} \cite{PhysRevX.7.031012}, where they suggest and demonstrate pulse shaping of photon subtractors by means of gate pulses and frequency up-conversion. In their method, however, higher order sideband modes have complex pulse shapes so as to achieve orthogonality, and it gradually gets difficult to actually use such higher frequency modes.

In conclusion, we have proposed and experimentally realized a highly pure photon subtractor that operates on high-frequency sideband modes of light. The target DSB mode is suitable to the frequency-division multiplexing of non-Gaussian states. Our scheme is applied to the creation of an SC state on a 500\,MHz sideband with about 5\,MHz of bandwidth and nagativity in the Wigner function is observed.
Our techniques developed here can be applied to higher order sidebands of the OPO just by changing the frequency of the phase modulation, while keeping the time-domain shape of the envelope. With a carrier frequency LO, any DSB modes at various frequencies can be simultaneously measured in principle \cite{PhysRevA.77.063817}. In addition to such multi-frequency encoding, it is notable that two quadrature sideband modes (sin and cos) at one frequency are also useful for dual-rail encoding of quantum states. Since the DSB encoding (cos- and sin-sideband modes) and single-sideband encoding (upper- and lower-sideband modes) are connected by effective beamsplitter transformations, these encoding can be used for a single-beam implementations of quantum teleportation \cite{PhysRevA.90.042337} or cat breeding protocols \cite{sychev2017enlargement}.

This work was supported by CREST (JPMJCR15N5) of JST, JSPS KAKENHI, and the Australian Research Council Centre of Excellence for Quantum Computation and Communication Technology (Project No. CE170100012).

%

\clearpage
\onecolumngrid
{\centering \large Supplemental Material for\\}
{\centering \Large Schr\"odinger's cat in an optical sideband\\}
\vspace{8pt}
{\centering Takahiro Serikawa, Jun-ichi Yoshikawa, Shuntaro Takeda, Hidehiro Yonezawa,\\ Timothy C. Ralph, Elanor H. Huntington, and Akira Furusawa\\}
\vspace{18pt}
\twocolumngrid

{\bf Model of sideband squeezing.} -- 
The linewidth of an optical parametric oscillator (OPO) induces finite time correlation of the output squeezed state. In this section, we describe the broadband squeezing operation induced by an OPO to deduce the squeezing spectrum and time correlation of sideband squeezed states. The unit of $\hbar=1$ is adopted throughout this supplementary material.
The squeezing operation with squeezing spectrum $r(\omega)$ is expressed as
\begin{align}
 |\psi\rangle_\mathrm{OPO} = \hat{S}^{{\cos}\dagger}_r\hat{S}^{{\sin}\dagger}_r|0\rangle,
 \label{eq:opoepr}
\end{align}
where $\hat{S}_r^i = \exp\bigl(\int_0^\infty d\omega\, r(\omega) (\hat{a}^{i\dagger}_\omega)^2/2 - \mathrm{h.c.}\bigr)$, $i = {\cos}, {\sin}$ is the broadband squeezing operator acting on the DSB basis. This is equivalent to Bogoliubov transformation of each sideband mode expressed in the Heisenberg picture as
\begin{align}
 \hat{S}^{i\dagger}_r\hat{a}^i_\omega\hat{S}^i_r = \cosh|r(\omega)|\, \hat{a}^i_\omega + e^{i\mathrm{Arg}\, r(\omega)} \sinh|r(\omega)|\, \hat{a}^{i\dagger}_\omega.
 \label{eq:sidebandbogtrans}
\end{align}
The quadrature operator of a sideband mode along the optical phase $\phi$ is defined as
\begin{align}
 \hat{x}_\omega^i(\phi) = \frac{1}{2}\left(e^{i\phi}\hat{a}^i_\omega + e^{-i\phi}\hat{a}^{i\dagger}_\omega\right).
\end{align}
Anti-squeezing / squeezing is realized at $2\phi + \mathrm{Arg}\, r(\omega) = 0,\pi/2$ respectively and then the quadrature variances $V^i(\omega; \phi) = \langle \hat{x}^i_\omega(\phi)^2\rangle$ read
\begin{align}
 V^i\Bigl(\omega; -\mathrm{Arg}\,r(\omega)/2\Bigr) = \frac{1}{2}e^{2|r(\omega)|},
\end{align}
\begin{align}
 V^i\Bigl(\omega; \pi/4-\mathrm{Arg}\,r(\omega)/2\Bigr) = \frac{1}{2}e^{-2|r(\omega)|}.
\end{align}
The general form of the squeezing level spectrum $r(\omega)$ is given in reference \cite{PhysRevA.73.013817}. When we assume the frequency structure of the OPO to be symmetric about $\omega_0$ and the pump phase to be zero, $r(\omega)$ is real regardless of $\omega$, giving
\begin{align}
 r(\omega) = \ln\left|\frac{(\gamma + \epsilon)^2 - \left(\frac{1 \pm e^{i\omega \delta}}{\delta}\right)^2}{\left(\gamma - \frac{1 \pm e^{i\omega \delta}}{\delta}\right)^2 - \epsilon^2}\right|,
\end{align}
where $\gamma$ is the cavity decay constant, $\epsilon$ is the pump parameter, and $\delta$ is the round-trip time of the cavity. Periodic structure appears in $e^{i\omega \delta}$ suggesting $2\pi/\delta$ as the free spectral range (FSR) of the OPO, which we define as $2\Omega$. In this formula, $\pm$ corresponds to the two possibility of the OPO resonance condition, i.e, $-$ for when the OPO is resonant at $\omega=0$ and $+$ for when the OPO is anti-resonant. Here, we consider the latter case and focus on the lowest-frequency resonance. The squeezing spectrum is approximated around $\omega = \Omega$, leading to
\begin{align}
 r(\omega) \approx \ln\left|\frac{\gamma + \epsilon + i(\omega - \Omega)}{\gamma - \epsilon - i(\omega - \Omega)}\right|.
 \label{eq:sidebandsqspec}
\end{align}
This form is justified when each resonant peak of the OPO is narrow and well separated from each other, as is the case in our experiment. When the pump field is weak, Eq. (\ref{eq:sidebandsqspec}) is further approximated,
\begin{align}
 r(\omega) \approx \frac{2\gamma\epsilon}{\gamma^2 + (\omega - \Omega)^2},
 \label{eq:sidebandspecapprox}
\end{align}
giving a Lorentzian spectrum around the resonance at $\Omega$. Considering the detection efficiency $\eta$, the squeezing level $R_\pm(\omega)$ is expressed as
\begin{align}
 R_\pm(\omega) = 1 + \eta \left[\exp\frac{4\gamma\epsilon}{\gamma^2 + (\omega - \Omega)^2} - 1\right].
\end{align}

These frequency-domain analyses are exported to a time-domain by introducing instantaneous mode operators
\begin{align}
 \hat{a}_t = \frac{1}{\sqrt{2\pi}}\int d\omega\, e^{i\omega t} \hat{a}_\omega.
\end{align}
Sideband modes are sine and cosine transforms of $\hat{a}_t$;
\begin{align}
 \hat{a}^\mathrm{cos}_\omega = \frac{1}{\sqrt{2\pi}}\int dt\, \cos\omega t\, \hat{a}_t,\\
  \hat{a}^\mathrm{sin}_\omega = \frac{1}{\sqrt{2\pi}}\int dt\, \sin\omega t\, \hat{a}_t.
 \label{eq:sincostrans}
\end{align}
The inverse transformation of Eq. (\ref{eq:sincostrans}) is
\begin{align}
\hat{a}_t = \frac{1}{\sqrt{2\pi}}\int d\omega\, \Bigl(\cos\omega t\, \hat{a}^\mathrm{cos}_\omega + \sin\omega t\, \hat{a}_\omega^\mathrm{sin} \Bigr).
\label{eq:invcossin}
\end{align}
Equation (\ref{eq:opoepr}) is transformed as
\begin{align}
 |\psi\rangle_\mathrm{OPO} =  \exp\left[\int dt_1dt_2\, \Bigl(R(t_1-t_2) \hat{a}^\dagger_{t_1}\hat{a}^\dagger_{t_2} - \mathrm{h.c.} \Bigr)\right] |0\rangle,
\label{eq:opoeprtime}
\end{align}
where the time-domain correlation function $R(t)$ is defined by
\begin{align}
 R(t) = \frac{1}{\sqrt{2\pi}}\int d\omega\, e^{-i\omega t} r(\omega).
\end{align}
When we place the assumptions above, $R(t)$ is derived from Eq. (\ref{eq:sidebandspecapprox}) as
\begin{align}
 R(t) = \sqrt{2\pi} \epsilon \exp\left(-\gamma|t| - i\Omega t\right).
 \label{eq:timecol}
\end{align}
By defining the double-sided decay function of the OPO as $h(t) = \exp(-\gamma|t|)$, the time correlation of the first sideband squeezing is considered to be a product of cavity decay $h(t)$ and sideband-frequency rotation $\exp(-i\Omega t)$.

{\bf Model of sideband photon subtraction.} --
In this section, we describe photon subtraction from a DSB mode and analyze the effect on the sideband squeezed state. We show that optical Schr\"odinger's cat (SC) states are generated on wavepackets, whose envelope is determined by the response of frequency filters and the OPO's time correlation.

To clarify the merit of the proposed method, we first consider a simpler way of photon subtraction from sideband, where a frequency filter picks up the upper and lower band component of a sideband squeezed state and a photon detection at these bands heralds a subtraction event. This is a {\it passive} process and does not discriminate cos- and sin-sidebands, as a result of the time translation symmetry of the setup. Our scheme, in contrast, uses an {\it active} process, where the phase modulation selects one fixed DSB mode synchronized with an external reference frame.

\begin{figure}[t]
 \centering
 \includegraphics{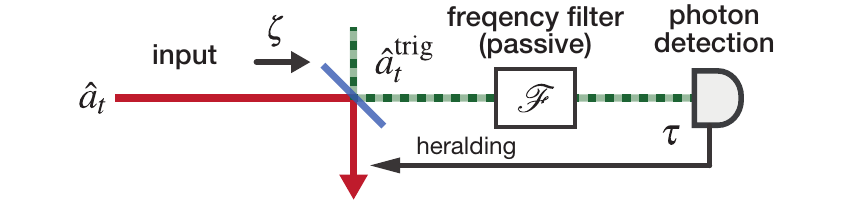}
 \caption{\label{fig:normalsubt} Passive  photon subtraction from double-sideband. A beamsplitter with the transmissivity $\zeta$ picks up trigger photons. Sideband photons are selected by a frequency filter $\mathcal{F}$ and detected.}
\end{figure}

Figure \ref{fig:normalsubt} shows a rough sketch of passive subtraction from a DSB mode. A small portion $\zeta$ of the light is picked up to the trigger mode $\hat{a}_t^\mathrm{trig}$ by a beamsplitter operation $\hat{B}$:
\begin{align}
 \hat{B}^\dagger \hat{a}_t^\mathrm{trig}\hat{B} = \sqrt{1-\zeta}\hat{a}^\mathrm{trig}_t + \sqrt{\zeta}\hat{a}_t.
 \label{eq:tappingbs}
\end{align}
The frequency filter $\mathcal{F}$ is placed to extract the first sideband component around $\pm\Omega$. We use Eq. (\ref{eq:sidebandspecapprox}) for the expression of the sideband squeezing, where the higher order resonances of the OPO are ignored since they are filtered out. Subsequently a photon at $t=\tau$ is detected, providing the conditioned state
\begin{align}
 |\psi\rangle_\tau = \mathcal{N}\left[{}_\mathrm{trig}\langle0|\hat{a}^\mathrm{trig}_\tau \hat{B}|\psi\rangle_\mathrm{OPO}|0\rangle_\mathrm{trig}\right],
\end{align}
where $\mathcal{N}$ is renormalization. Considering Eq. (\ref{eq:tappingbs}), this results in
\begin{align}
 |\psi\rangle_\tau = \mathcal{N}\left[{}_\mathrm{trig}\langle0|\hat{B}\Bigl(\hat{a}_\tau|\psi\rangle_\mathrm{OPO}\Bigr)|0\rangle_\mathrm{trig}\right].
\end{align}
Thus the conditioned state is a photon subtracted state with an optical loss of $\zeta$ induced by $\hat{B}$. From Eqs. (\ref{eq:opoepr}) and (\ref{eq:invcossin}),
\begin{align}
\begin{aligned}
 &\hat{a}_\tau |\psi\rangle_\mathrm{OPO} \propto\\
 & \int d\omega\, \Bigl(\cos\omega\tau\, \hat{a}^\mathrm{cos}_\omega + \sin \omega\tau\, \hat{a}_\omega^\mathrm{sin}\Bigr)\left(\hat{S}^{\cos}_r \hat{S}^{\sin}_r |0\rangle\right).
\end{aligned}
\end{align}
Here we set $\tau = 0$, in other words, the origin of sideband phase is re-defined by the photon detection timing. Then the photon is subtracted from cos-sideband and the two-mode states are factorized in the DSB basis.
\begin{align}
 \hat{a}_0 |\psi\rangle_\mathrm{OPO} \propto \left( \int d\omega\, \hat{a}^\mathrm{cos}_\omega \hat{S}^{\cos}_r |0\rangle_{\cos} \right) \otimes \left(\hat{S}^{\sin}_r |0\rangle_{\sin}\right).
\end{align}
Now we show that the photon-subtracted cos-sideband mode is in a squeezed single photon state, i.e. an SC state \cite{PhysRevA.55.3184}. Equations (\ref{eq:sidebandbogtrans}) and (\ref{eq:sidebandsqspec}) lead to
\begin{align}
 \int d\omega\, \hat{a}^\mathrm{cos}_\omega \hat{S}^{\cos}_r |0\rangle_{\cos} = \hat{S}^{\cos}_r \int d\omega\, \sinh r(\omega)\,\hat{a}^{\dagger\mathrm{cos}}_\omega|0\rangle_{\cos},
\end{align}
where $\int d\omega\, \sinh r(\omega)\,\hat{a}^{\dagger\mathrm{cos}}_\omega|0\rangle_{\cos}$ expresses a single photon state in cos-sideband wavepacket. In the weak pumping regime, $\sinh r(\omega)$ is near to $r(\omega)$ and we can assume Eq. (\ref{eq:sidebandspecapprox}), resulting in the time-domain expression of the single photon state wavepacket:
\begin{align}
 \int d\omega\, \sinh r(\omega)\,\hat{a}^{\dagger\mathrm{cos}}_\omega|0\rangle = \int dt\, \exp(-\gamma|t|) \cos \Omega t \, \hat{a}_t^\dagger |0\rangle,
 \label{eq:directsubwavepacket}
\end{align}
where the envelope is $h(t)$ and carrier wave is $\cos \Omega t$. Therefore, the conditioned state can be interpreted as a squeezed single-photon state in the cos-sideband wavepacket $\hat{a}^\mathrm{passive} = \int dt\, h(t)\cos\Omega t\, \hat{a}_t$ and all the other modes are left in the sideband-squeezed state.

Although sideband-cat state can be generated by this method, it can be pointed out that the high frequency sideband is technically challenging to reach for the following reasons; 1. The timing jitter of trigger line electronics mixes sin and cos sideband components. Especially the timing resolution of state-of-the-art APDs, for example 225\,ps for SPCM-AQRH-TR series (Excelitas Technologies), will limit the sideband frequency at several hundred MHz. 2. The sideband phase is determined by the detection timing of the trigger photon. Such a posteriori phase makes it difficult to apply successive modulations or demodulations on the generated state or to make interference between multiple sideband states. 3. The trigger line frequency filter is hard to be realized by simple cavity filters since it should have a strenuous response such that $\pm\Omega$ is selected and $\pm3\Omega$, $\pm5\Omega$ and higher order resonances are rejected.

Now we move on to the active subtraction illustrated in the main text. The EOM picks up trigger photons from a certain sideband, which fixes the sideband phase of the heralded state irrespective of the photon detection timing. Phase modulation is expressed in the Heisenberg picture as \cite{Capmany:10}
\begin{align}
 \hat{a}^\mathrm{mod}_t &= \exp\left[i\beta \sin(\Omega t + \theta)\right]\hat{a}_t \\
  &= \sum_{n = -\infty}^\infty J_n(\beta) e^{in(\Omega t + \theta)} \hat{a}_t
  \label{eq:modtimedomain}
\end{align}
where $J_n$ is $n$-th order first kind Bessel function and $\theta$ is modulation phase. The frequency-domain expression of Eq. (\ref{eq:modtimedomain}) becomes
\begin{align}
 \hat{a}^\mathrm{mod}_\omega = J_0(\beta)\,\hat{a}_\omega + \sum_{n=1}^\infty J_n(\beta)\, \left[e^{in\theta} \hat{a}_{\omega+n\Omega} + e^{-in\theta} \hat{a}_{\omega - n\Omega}\right].
 \label{eq:modinfreq}
\end{align}
When $\beta \ll 1$, higher order terms can be omitted, giving
\begin{align}
 \hat{a}^\mathrm{mod}_\omega \approx \sqrt{1 - \frac{\beta^2}{2}} \hat{a}_\omega +\frac{\beta}{\sqrt{2}} \frac{e^{i\theta}\,\hat{a}_{\omega + \Omega} + e^{-i\theta}\, \hat{a}_{\omega-\Omega}}{\sqrt{2}}.
 \label{eq:smallmodfreq}
\end{align}
This is a sideband-phase-sensitive beamsplitter, whose sideband phase can be tuned by $\theta$. From now on, $\theta$ is set at zero. The baseband mode is coupled to the cos-sideband as
\begin{align}
 \hat{a}^\mathrm{mod}_{\omega = 0} \approx \sqrt{1 - \frac{\beta^2}{2}} \hat{a}_{\omega=0} +\frac{\beta}{\sqrt{2}} \hat{a}^\mathrm{cos}_\Omega.
 \label{eq:modcarrier}
\end{align}
Cos-sideband modes couple to both baseband and $2\Omega$ modes while the baseband component vanishes in the sin-sideband,
\begin{align}
 \hat{a}^{\cos,\, \mathrm{mod}}_{\Omega} &\approx \sqrt{1 - \frac{3\beta^2}{4}} \hat{a}^{\cos}_{\Omega} +\frac{\beta}{\sqrt{2}} \left[\hat{a}_{\omega=0} + \frac{1}{\sqrt{2}}\hat{a}^{\cos}_{2\Omega}\right] \\
 \hat{a}^{\sin,\, \mathrm{mod}}_{\Omega} &\approx \sqrt{1 - \frac{\beta^2}{4}} \hat{a}^{\sin}_{\Omega} +\frac{\beta}{2} \hat{a}^{\sin}_{2\Omega}.
 \label{eq:phasemodsideband}
\end{align}
In the time-domain, the assumption $\beta \ll 1$ leads to
\begin{align}
  \hat{a}_t^\mathrm{mod} \approx \left[\sqrt{1 - \frac{\beta^2}{2}} + \beta\cos\Omega t\right] \hat{a}_t.
 \label{eq:smallmodtime}
\end{align}

Now we consider sideband photon subtraction. The OPO's output mode $\hat{a}_t$ is transformed to $\hat{a}^\mathrm{mod}_t$ by a phase modulation at the first sideband frequency $\Omega$. The series of a frequency separator and filters are in total an optical low-pass filter represented by a time-domain response function $f(t)$. The filtered mode $\hat{a}_t^\mathrm{trig}$ is expressed as a sum of transmission response of $\hat{a}^\mathrm{mod}_t$ and reflection response of $\hat{a}_t^\mathrm{trig}$:
\begin{align}
 \hat{a}_t^\mathrm{trig} = \int dt'\,\left[ f(t'-t)\, \hat{a}^\mathrm{mod}_{t'} + \Bigl(\delta(t) - f(t'-t)\Bigr)\hat{a}^\mathrm{trig}_{t'}\right],
\end{align}
where the initial state of $\hat{a}_t^\mathrm{trig}$ is assumed to be vacuum. Substituting $\hat{a}_t^\mathrm{mod}$ with Eq. (\ref{eq:smallmodtime}), $\hat{a}_t^\mathrm{trig}$ is split to three terms:
\begin{align}
\begin{aligned}
 \hat{a}_t^\mathrm{trig} = \int dt'\,\Biggl[\sqrt{1 - \frac{\beta^2}{2}}&f(t'-t)\, \hat{a}_{t'}  + \beta\cos\Omega t'\, f(t'-t)\, \hat{a}_{t'}  \\
 &+ \Bigl(\delta(t) - f(t'-t)\Bigr)\hat{a}^\mathrm{trig}_{t'}\Biggr].
\end{aligned}
\end{align}
We assume that the low-pass cutoff is below the first sideband frequency. This derives the following interpretations; the first term is composed of the baseband vacuum component of the OPO's output, and the second term is the downconverted component, which carries photons from the sideband squeezing. A photon detection at $t=\tau$ places ${}_\mathrm{trig}\langle 0|\hat{a}_\tau^{\mathrm{trig}}$ on the OPO's output state $|\psi\rangle_\mathrm{OPO}$. Here, since the only photon source in $\hat{a}_t^\mathrm{trig}$ is the second term, the other terms disappear. The conditioned state reads
\begin{align}
 |\psi\rangle_\tau = \int dt\, f(t-\tau)\, \cos\Omega t\, \hat{a}_{t}\,|\psi\rangle_\mathrm{OPO}.
\end{align}
Thus, a photon is subtracted from the following sideband-wavepacket mode:
\begin{align}
 \hat{a}^\mathrm{sub}_\tau = \int dt\, f(t-\tau) \cos \Omega t\, \hat{a}_t.
\end{align}
As for conventional baseband photon subtraction schemes, a squeezed single photon state, i.e. an optical SC state is induced by this operation. $\hat{a}^\mathrm{sub}_\tau$ and OPO's correlation function determines the SC state's mode $\hat{a}^\mathrm{cat}_\tau$ in a similar manner \cite{PhysRevA.96.052304} to Eq. (\ref{eq:directsubwavepacket}), giving
\begin{align}
 \hat{a}^\mathrm{cat}_\tau = \int dt\ (f*h)(t - \tau) \cos\Omega t\, \hat{a}_t,
 \label{eq:catmode}
\end{align}
where the envelope function is a convolution of filter response $f(t)$ and OPO's decay function $h(t)$, namely $(f*h)(t) = \int dt'\, f(t') h(t - t')$. This is the envelope of the cat state $\xi(t)$ in the main text. The orthogonal sideband mode
\begin{align}
 \hat{a}^\mathrm{SQ}_\tau = \int dt\ (f*h)(t - \tau) \sin \Omega t\, \hat{a}_t
\end{align}
remains in a squeezed state. Note that $\tau$ only appears in the envelope, while the preliminary definition of the sideband phase is given by the downconverting EOM as $\cos\Omega t$. This feature greatly helps us to apply further operations on the generated cat state, since linear operations, for example quantum teleportation  or universal squeezing \cite{PhysRevA.71.042308}, can be applied continuously on the cos-sideband.

{\bf Model of sideband homodyne measurement.} --
A time-resolved homodyne measurement gives a way to access multiple DSB modes independently and simultaneously. An ideal homodyne detection is a quadrature measurement of instantaneous mode $\hat{a}_t$:
\begin{align}
 \hat{X}(t; \phi) = \frac{e^{i\phi}\hat{a}^{\dagger}_t + e^{-i\phi}\hat{a}_t}{\sqrt{2}},
\end{align}
which contains the information of the quadrature of any longitudinal modes. Quadrature of a wavepacket mode with a real mode function $g(t)$ can be calculated as
\begin{align}
 \hat{X}_{g}(\phi) = \int dt\, g(t) \hat{X}(t; \phi).
\end{align}

\begin{figure*}[t]
 \centering
 \includegraphics{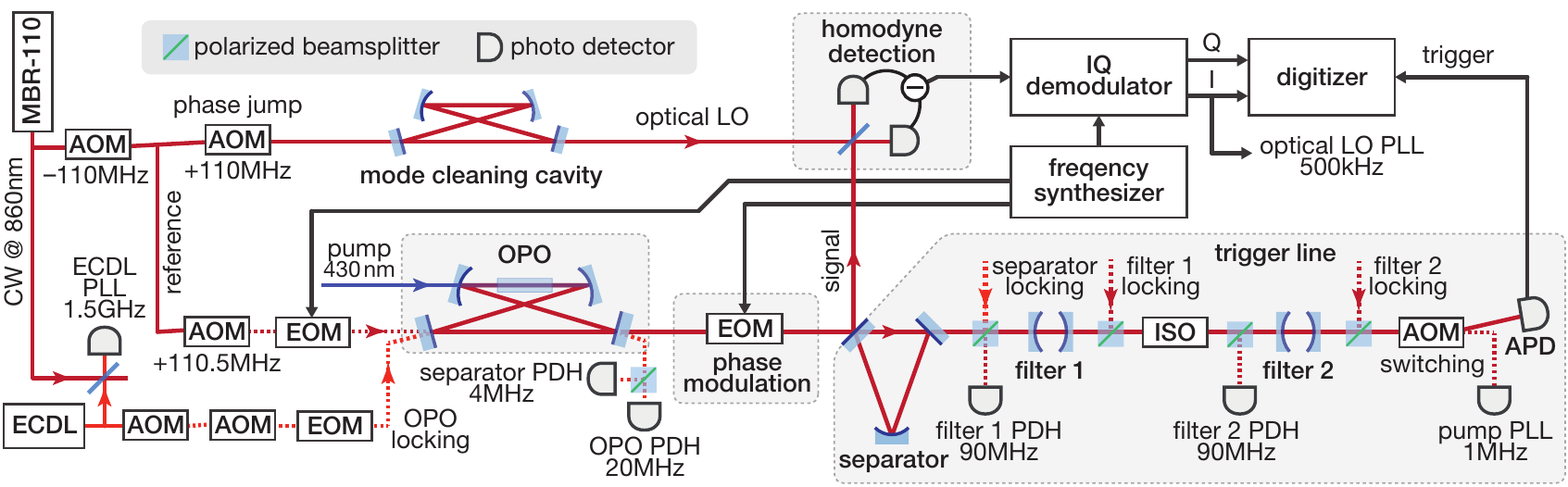}
 \caption{\label{fig:expdetail} Experimental setup. Solid lines are continuous wave laser and dashed lines indicates chopped light. CW: continuous-wave; AOM: acoust-optic modulator; ECDL: extra cavity diode laser; ISO: optical isolator; PDH: Pound-Drever-Hall locking, PLL: phase-locked loop.}
\end{figure*}

We first multiply the sideband envelope function electrically and continuously by an IQ demodulator. This results in two instantaneous signals
\begin{align}
  \hat{X}^{\cos}(t;\phi) &=\cos\Omega t\, \hat{X}(t; \phi), \\
 \hat{X}^{\sin}(t;\phi) &= \sin\Omega t\, \hat{X}(t; \phi).
\end{align}
In our experiment, the bandwidth of the homodyne detector around the sideband frequency is not negligible. Thus the detected signal is a convolution with a response function $d(t)$,
\begin{align}
 \hat{X}_d^{\cos}(t;\phi) = \int dt'\, d(t -t')\cos\Omega (t')\, \hat{X}(t'; \phi).
\end{align}
We assume that the response of the homodyne detector is determined by the first order resonance, which behaves as a single pole low-pass filter around the sideband frequency expressed by
\begin{align}
 d(t) = \exp(-2\pi f_c t) \Theta(t),
\end{align}
where $f_c$ is the cut-off frequency, and $\Theta(t)$ is Heaviside step function.

The waveform is digitized with triggers of photon detection events at $\tau$. From now on, we re-label $t$ so that the trigger time $\tau$ comes to the origin, thereby shifting the envelope of the wavepacket at the fixed position in the measurement frame. Then the data trace is expressed as
\begin{align}
 \hat{X}_d^{\cos}(t;\phi) = \int dt'\, d(t- t')\cos\Omega (t' + \tau)\, \hat{X}(t' + \tau; \phi).
\end{align} 
The wavepacket of the cat state is identified from the digitized data ensemble by applying independent component analysis (ICA), which detects the non-Gaussianity of the heralded cat mode $\hat{a}^\mathrm{cat}_\tau$. We optimize the envelope function $\chi(t)$ of the cat mode minimizing the kurtosis of the quadrature distribution. Thus we expect that the quadrature of the cat state $\hat{X}^\mathrm{cat}(\phi)$ is extracted as follows:
\begin{align}
 \hat{X}^\mathrm{cat}(\phi) &= \int dt\, \chi(t) \hat{X}^{\cos}_d(t;\phi)\\
  &= \int dt\, (f*h)(t) \cos\Omega (t + \tau)\, \hat{X}(t; \phi).
\end{align}
$\chi(t)$ is supposed to be a deconvolution of $d(t)$ from the naive envelope $(f*h)(t)$ in Eq. (\ref{eq:catmode}). As long as the detector's response $d(t)$ is narrow enough compared to $(f*h)(t)$, the estimated mode function $\chi(t)$ is non-singular, and the quadrature of the cat state can be numerically extracted. Using the same envelope $\chi(t)$, the orthogonal sideband mode is extracted as
\begin{align}
 \hat{X}^\mathrm{SQ}(\phi) &= \int dt\, \chi(t) \hat{X}^{\sin}_d(t;\phi)\\
  &=  \int dt\, (f*h)(t) \sin\Omega (t + \tau)\, \hat{X}(t; \phi).
\end{align}

\begin{figure}[t]
 \centering
 \includegraphics{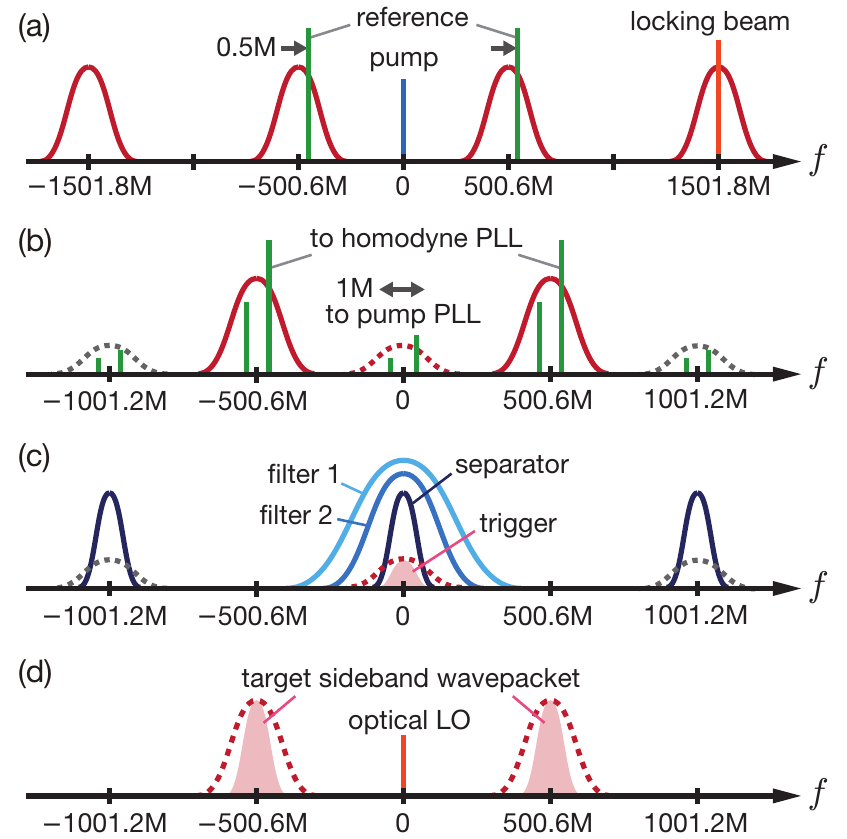}
 \caption{\label{fig:spec} Frequency diagram of the experiment. Frequency $f$ is expressed in Hz. (a) OPO's resonance, and pump / locking / reference beam. (b) Spectrum of the squeezed light and the reference beam after going through the OPO and downconverting EOM. (c) Filtering spectrum of the trigger line. The resonance of the separator cavity and two filtering cavity is indicated. (d) The frequency structure of the homodyne detection. The SC state is generated in a narrower bandwidth than the resonance of the OPO.}
\end{figure}

{\bf Experiment.} --
Figure \ref{fig:expdetail} shows the detail of the experiment. The OPO with an FSR of $2\Omega=1001.2\,\mathrm{MHz}$ includes a type-0 phase matched PPKTP crystal (1\,mm $\times$ 1\,mm $\times$ 10\,mm, Raicol), which operates a degenerate parametric down conversion at least over 100\,GHz bandwidth. The p-polarized pump light at 430\,nm is provided by a second-harmonic-generator cavity driven by an 860\,nm continuous-wave seed laser (MBR-110, Coherent). The oscillation threshold of the OPO is measured at 550\,mW and the power of the pump light is set at 25\,mW, corresponding to the normalized pump amplitude $\epsilon = 0.21$. The OPO's output coupler mirror with 12\% transmissivity corresponds to the line width of $f_\mathrm{HWHM} = 10\,\mathrm{MHz}$, or equivalently, the decay constant $\gamma_\mathrm{OPO} = 1/16\,\mathrm{ns}$. The design parameters of the cavities are summarized in Table \ref{fig:linewidth}. The OPO is locked by a $1501.8\,\mathrm{MHz}$ detuned locking beam produced by an external cavity diode laser to make it resonant at $\pm \Omega, \pm3\Omega, \cdots$ sidebands (Fig. \ref{fig:spec}).

\begin{table}[b]
 \centering
 \begin{tabular}{cccc}
   & $f_\mathrm{FSR}$ & $f_\mathrm{HWHM}$ & $\gamma$\\ \hline 
  OPO & 1.0012\,GHz & 10\,MHz & 1/16\,ns\\
  separator & 1.00\,GHz & 5.3\,MHz & 1/30\,ns\\
  filter 1 & 75\,GHz & 72\,MHz & 1/2.2\,ns\\
  filter 2 & 52\,GHz & 50\,MHz & 1/3.2\,ns\\ \hline
 \end{tabular}
 \caption{\label{fig:linewidth} Parameters of the OPO and the trigger line filters. $f_\mathrm{FSR}$: free spectral range, $f_\mathrm{HWHM}$: linewidth in half-width at half-maximum (HWHM), $\gamma$: decay constant.}
\end{table}

The output squeezed light is modulated by a homemade EOM, which consists of a bulk KTP crystal (1\,mm $\times$ 1\,mm $\times$ 10\,mm, Raicol) and air-core transformer for the resonant matching at 500\,MHz. The modulation power is adjusted at 29\,dBm so that 2.0\% of the light power in 500.6\,MHz cos-sideband is downconverted to 0\,Hz, corresponding to $\beta^2 = 0.040$ in Eq (\ref{eq:modcarrier}). As in Eq. (\ref{eq:modinfreq}), part of the optical power is also distributed to the second or higher order harmonic, however, the transfer ratio is below 0.1\% and negligible. A high-grade anti-reflection coating suppresses the optical loss below 0.5\%. The sideband squeezed light and the downconverted trigger photons are subsequently split by a triangle cavity. The sideband signal light is reflected and measured by homodyne detection with 0\,Hz optical local oscillator (LO). The LO beam is spatially and longitudinally filtered by a mode cleaning cavity. We have developed a resonant homodyne detector for a direct detection of 500\,MHz sideband, which is equipped with 98\% quantum efficiency Si photodiode (S5971SPL, Hamamatsu Photonics) and has 12.0\,dB of shotnoise signal to noise ratio at 500\,MHz with 5.0\,mW LO \cite{doi:10.1063/1.5029859}. The cut-off frequency $f_c$ is measured at 14\,MHz by fitting the gain spectrum with Lorentzian function around the resonance peak. Since the electric noise can be equivalently treated as an optical loss \cite{PhysRevA.75.035802}, we derive a frequency-dependent loss spectrum at below 7\% across the OPO's bandwidth. The IQ demodulator (ADL5380, Analog Devices) driven by 500.6\,MHz electrical LO downconverts in-phase (cos-sideband) and quadrature (sin-sideband) components of the homodyne detector's signal. They are simultaneously digitized by a 5\,Gsamples/s, 12-bit oscilloscope (DSOS204A, Keysight Technologies) with a trigger signal from the APD.

The overall optical efficiency of the setup is estimated as $\eta_\mathrm{tot}^\mathrm{est}=0.80$, which includes the OPO's escape efficiency (0.982), the propagation loss (0.035), the interference efficiency with optical LO (0.935), and the detection efficiency of the homodyne measurement (0.91). From Eq. (\ref{eq:phasemodsideband}), the power transfer ratio at the phase modulation, $3\beta^2/4 = 0.030$ for cos-sideband and $\beta^2/4 = 0.010$ for sin-sideband, is also to be taken into account.
From the squeezing level shown in the main text, the detection efficiency of sin-sideaband $\eta^{\sin}$ is measured at 70\%.
Deducting the 1\% power transfer by the phase modulation at sin-sideband, we derive the total efficiency $\eta_\mathrm{tot}$ at 71\% and estimate the effective efficiency of cos-sideband $\eta^{\cos}$ at 68\%. We have 9\% of excessive loss in $\eta_\mathrm{tot}$ over $\eta_\mathrm{tot}^\mathrm{est}$. We currently do not identify this difference, while we suppose it to be an optical loss of photodiode in the high-frequency region. Thus we assume the actual detection efficiency of the homodyne measurement $\eta_\mathrm{det}$ at 83\%, as mentioned in the main text.

The trigger photon is further filtered by two Fabry-Perot cavities to reject higher order sideband photons. The response function $\mathcal{T}(t)$ of each filter is exponential decay function characterized by the decay constant $\gamma$ shown in Table S1:
\begin{align}
 \mathcal{T}(t) = \exp(-\gamma t)\Theta(t).
\end{align}
The total response function is a convolution of the three responses. All these filtering cavities are locked by Pound-Drever-Hall method using s-polarized counter-propagating locking beams. The trigger photon is detected by an APD (SPCM-AQRH-16-FC, Excelitas Technologies), supplying a trigger signal which heralds the photon-subtraction event. The total transmission of the trigger line is 39\%, and the total detection efficiency is 10\% considering the APD's quantum efficiency of 54\% and 50\% intrinsic loss at the picking-up modulation.

To control optical phase and sideband phase, a p-polarized reference light is introduced to the OPO through a high-reflection mirror. This light is detuned at 500\,kHz and deeply modulated at $\Omega=500.6\,\mathrm{MHz}$ by a waveguide-type phase modulator (EOSPACE) to go into the $\pm500$\,MHz resonance of the OPO (Fig. \ref{fig:spec}). The relative phase of the downconverting EOM and the electrical LO of the IQ demodulator is adjusted so that they match the modulation phase of the reference light. All these modulation signals are generated from the synchronized direct digital synthesizers (AD9959, Analog Devices). When $0.5\pm500.6$\,MHz sideband light goes through the OPO, the parametric amplification generates a difference frequency light at $-0.5\pm500.6$\,MHz. After the EOM, 1\,MHz amplitude modulation appears, to which the phase-locked loop (PLL) technique is applied to lock the pumping phase. The optical LO phase is also controlled by PLLs using the 500\,kHz signal of the in-phase component of the homodyne detection. The reference beam and cavity-locking beams are chopped during the measurement period and the phase locking are held then. The interval of the sample-and-hold loop is 250\,us and the duration of the measurement window is 60\,us. To scan the phase of the homodyne measurement, we use an acoust-optic modulator (AOM) to jump the LO phase from the constant locked phase to the arbitrary measurement phase at the beginning of the window. 

The event rate of the photon subtraction is 900\,counts/s inside the measurement window. Among the triggers, 7 counts/s is fake clicks, which is mainly the dark count of the APD. This replaces 0.8\% of the triggered states with non-heralded squeezed states.

\begin{figure}[t]
 \centering
 \includegraphics{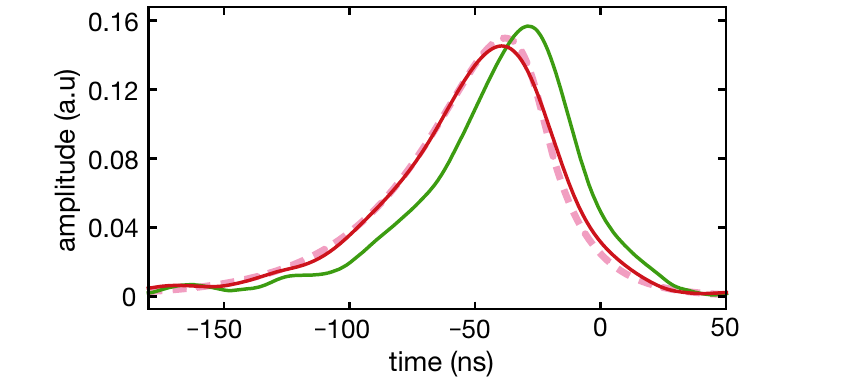}
 \caption{\label{fig:wavepacket_analysis} Envelope functions of the sideband cat state. Green curve: analytical envelope function $\chi(t)$ estimated by ICA. Red curve: physically measured temporal mode envelope $\xi(t)$, calculated as $(\chi*d)(t)$; and blue-dashed curve: theoretical prediction of the envelope function of the cat state $(f*h)(t)$. These two are shown in the main text. All these wavepackets are normalized by 2-norm.}
\end{figure}

{\bf Mode estimation and state tomography.} --
8000 traces of the demodulated homodyne signal $X_d^{\cos}(t;\phi)$ and $X_d^{\sin}(t;\phi)$ are collected for equally separated 36 measurement phases. Figure \ref{fig:wavepacket_analysis} shows the envelope function of the photon-subtracted mode $\chi(t)$ obtained from the cos-sideband dataset by ICA. The physically measured envelope with the homodyne detector $(\chi * d)(t)$ fits the theoretical envelope function $(f*h)(t)$ that we expected, showing 99\% of mode-matching. The quadrature distributions are obtained by multiplying $\chi(t)$ to the datasets. We normalize them with the shotnoise variance of the same wavepacket, which is acquired from the homodyne detection without pumping the OPO.

\begin{figure}[b]
 \centering
 \includegraphics{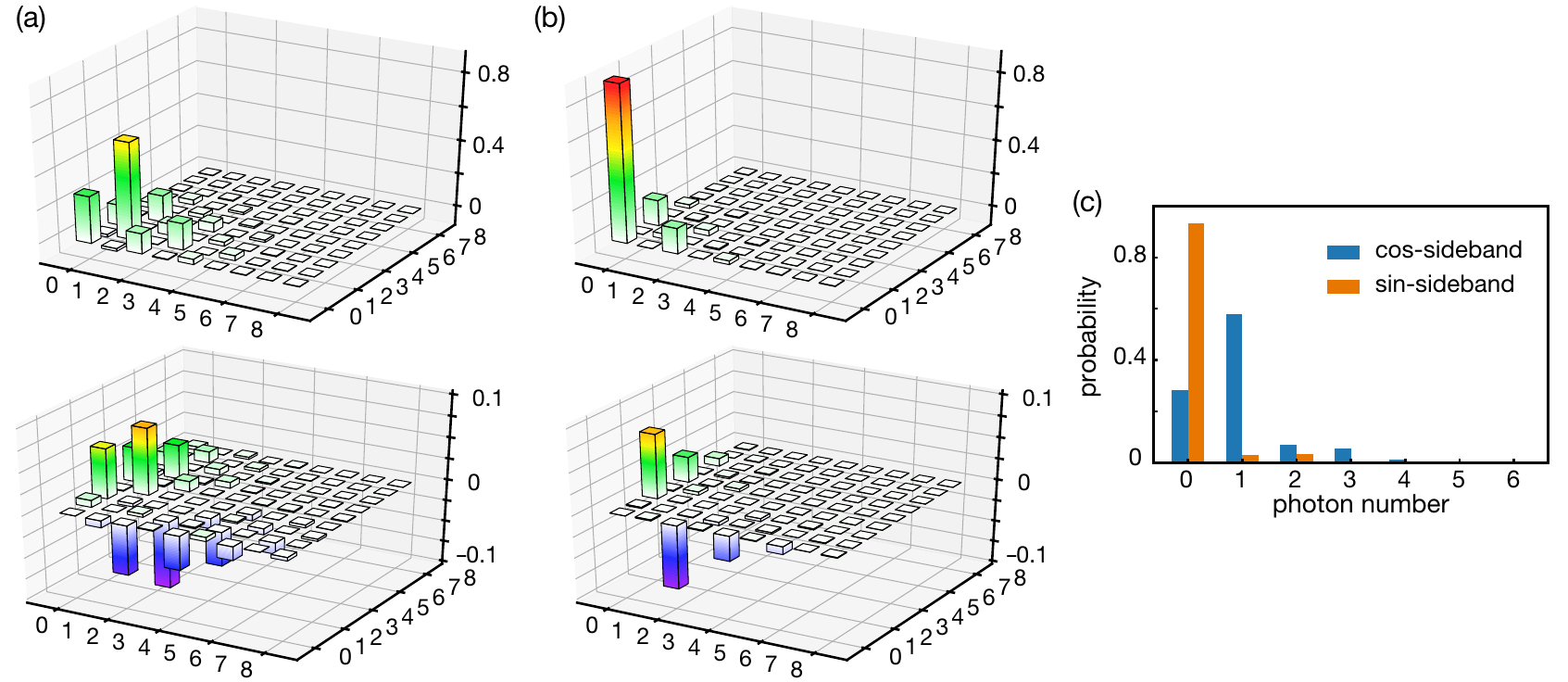}
 \caption{\label{fig:densitymatrix} Density matrices of (a) cos-sideband mode and (b) sin-sideband mode. The real part (above) and imaginary part (below) are separately plotted. Only the subspace up to 8 photons is shown. (c) photon number distributions.}
\end{figure}

Recursive maximum-likelihood estimation \cite{lvovsky2004iterative} is used to reconstruct the density matrix in Fock basis with a photon-number cutoff of 13. Figure \ref{fig:densitymatrix} shows the density matrices estimated from the quadrature distributions of cos- and sin- sideband. Wigner function representation in the main text is obtained from the density matrix.
To estimate the statistical error of the Wigner function, we use bootstrap method \cite{efron1986bootstrap}, where the variance of the Wigner function is calculated from 100 times of trial virtually performed by resampling 8000 quadratures from the raw data for each phase.

\onecolumngrid

\end{document}